\documentclass[prl,10pt,twocolumn,superscriptaddress,showpacs]{revtex4-1}

\usepackage{epsfig}
\usepackage{amsmath}
\usepackage{graphicx}
\usepackage{graphics}
\usepackage{float}
\usepackage{amssymb}
\usepackage[usenames,dvipsnames]{color}
\usepackage{natbib}
\usepackage{epstopdf}
\usepackage{verbatim}
\usepackage{wrapfig}

\newcommand{\ket}[1]{\left| #1\right\rangle}

\newcommand{\etal}{{\it et al.}}
\newcommand{\jn}[1]{~{\bf #1}}

\begin{document}

\title{Protecting and Enhancing Spin Squeezing via Continuous Dynamical Decoupling}

\author{Adam Zaman Chaudhry}
\affiliation{NUS Graduate School for Integrative Sciences
and Engineering, Singapore 117597, Singapore}
\author{Jiangbin Gong}
\email{phygj@nus.edu.sg}
\affiliation{NUS Graduate School for Integrative Sciences
and Engineering, Singapore 117597, Singapore}
\affiliation{Department of Physics and Center for Computational
Science and Engineering, National University of Singapore, Singapore 117542,
Singapore}

\begin{abstract}
Realizing useful quantum operations with high fidelity is a two-task quantum control problem wherein
decoherence is to be suppressed and desired unitary evolution is to be executed.
The dynamical decoupling (DD) approach to decoherence suppression has been fruitful
but synthesizing DD fields with certain quantum control fields may be experimentally demanding.
In the context of spin squeezing, here we explore
an unforeseen possibility that continuous DD fields may
serve dual purposes at once. In particular, it is shown that
a rather simple configuration of DD fields
can suppress collective decoherence and yield a $1/N$ scaling of the squeezing performance ($N$ is the number of spins),
thus making spin squeezing more robust to noise and much closer to the so-called Heisenberg limit.  The theoretical predictions
should be within the reach of current spin squeezing experiments.

\end{abstract}
\maketitle

The feasibility of ``dynamical decoupling" (DD) \cite{LloydPRA1998}
in effectively isolating a quantum system from its environment has attracted great theoretical and experimental interests.
Remarkable progress towards efficient protection of quantum states has been achieved \cite{UhrigPRL2007,LiuFrontphys,BiercukNature2009,Du2009}.
In contrast, high-fidelity protection of quantum operations (such as quantum gates or quantum metrology schemes)
is more challenging experimentally.
As decoherence must be suppressed during quantum operations, it is natural to synthesize DD
fields with other fields implementing a desired quantum operation.
However, this bottom-up approach may require complicated coherent control fields.
For example, explicit solutions to dynamically corrected quantum gates are sophisticated \cite{LloydPRL1999,ViolaPRL2009,ChaudhryPRA2012},
and even a simple quantum metrology protocol, when combined with DD, already becomes a rather involving practice \cite{RongEPL2011}.

A top-down approach to the protection of useful quantum operations should be a worthy direction, along which we aim to
better exploit the system's own Hamiltonian under DD fields.
In essence we are faced with a two-task problem: decoherence is to be suppressed and
desired (almost) unitary evolution is to be executed.  Is it possible to directly construct DD fields
serving the dual tasks at once?  Motivated by recent studies of decoherence effects on spin squeezing \cite{SimonPRA2002,LukinPRA2002,StocktonPRA2003,WangPRA2010,SinatraPRL2008,SinatraPRL2011,KurizkiPRL2011,WatanabePRA2012} and by
recent exciting experiments of spin squeezing \cite{GrossNature2010,RiedelNature2010}, here we use the spin squeezing
context to give a positive answer to our question. That is, it is feasible
to protect and enhance spin squeezing at the same time by searching for a special configuration of
DD fields. The dual roles of DD fields arise from two facts: (i) DD fields modulate both system-bath interaction and
the system Hamiltonian itself that describes the spin-spin interaction, and (ii) the system Hamiltonian itself under the modulation of DD fields
may generate more useful quantum evolution.
We emphasize that the enhancement in spin squeezing we achieve is not a secondary outcome
of decoherence suppression. Rather, the enhanced spin squeezing is far superior to what can be normally achieved
under ``decoherence-free" conditions. Indeed, we predict a $1/N$ scaling of the obtained spin squeezing performance,
where $N$ is the number of spins.  Our theoretical
results should be testable by modifying existing experiments.  In addition,
because spin squeezing is closely related to multi-partite entanglement \cite{TothPhysicsReports2009, AmicoRMP2008},
the results are also of interest to ongoing studies of entanglement protection \cite{Mukhtar10a,Mukhtar10b,Wang2011,Jiang2011,Wang2011PRA}.

Consider then a collection of $N$ identical spins (or qubits). In terms of the standard Pauli matrices,
the dynamics can be described by collective angular momentum operators $J_k = \frac{1}{2} \sum_{m = 1}^N \sigma_k^{(m)}$, with $k=x,y,z$.
In spin squeezed states, quantum fluctuations of the collective angular momentum in one direction are significantly reduced at the price of
increased uncertainty in another direction \cite{KitagawaUedaPRA1993, JianMaPhysicsReports2011}, thus offering higher
precision in quantum metrology \cite{WinelandPRA1994,footnoteprecision}. For instance, we can improve high-precision spectroscopy and atomic clocks which are currently limited by spin noise \cite{SantarelliPRL1999}.

 Using the angular momentum commutation relations, the two most widely used measures of spin squeezing, $\xi_S^2$ \cite{KitagawaUedaPRA1993} and $\xi_R^2$ \cite{WinelandPRA1994}, are found to be bounded by $1/N$ \cite{footnoteparametersexplanation}. This fundamental limit to the amount of spin squeezing achievable reflects the Heisenberg precision
 limit in quantum measurement \cite{JianMaPhysicsReports2011,LloydScience2004}.
 In practice, the achievable degree of squeezing is considerably worse than the $1/N$ limit for two main reasons.
 First, squeezing is in general degraded by decoherence or noise \cite{foot1}.  The environment tends to destroy squeezing, causing the sudden death of squeezing \cite{WangPRA2010}. It may also change the optimal squeezing time window in an unpredictable way, leading to a non-optimal squeezing generation \cite{RiedelNature2010}.  Second, Hamiltonians that can be implemented so far cannot reach the $1/N$ scaling in theory.
 For instance, in two recent experiments \cite{GrossNature2010,RiedelNature2010} based on two-mode Bose-Einstein condensate (BEC),
 the so-called one-axis twisting (OAT) Hamiltonian \cite{KitagawaUedaPRA1993} $H_{\text{OAT}} = \chi J_x^2$ is realized, which
 can at most generate $\xi_R^2\sim 1/N^{\frac{2}{3}}$, not to mention decoherence effects.
It is thus clear that protecting spin squeezing against decoherence and pushing spin squeezing towards the $1/N$ scaling
would be of wide interest.

We start by considering a OAT system interacting with an environment, with the total Hamiltonian modeled by
$H = H_0 + H_B + H_{\text{SB}}$, where, $H_0 = H_{\text{OAT}} = \chi J_x^2$, $H_B$ is the Hamiltonian of the environment, and $H_{\text{SB}}$ represents
the system-environment coupling.  $H_{\text{SB}}$  is assumed to be
\begin{equation}
\label{systemenvinteraction}
H_{\text{SB}} = B_x J_x + B_y J_y + B_z J_z,
\end{equation}
where the $B_k$ are arbitrary bath operators (or randomly fluctuating noise for a classical bath).
Though coupling terms that are nonlinear in $J_k$ are not considered here,
the $H_{\text{SB}}$ in Eq.~(\ref{systemenvinteraction}) is already quite general insofar as it covers a broad class of problems
with both dephasing and relaxation.


 There is a standard route to seek a control Hamiltonian $H_c(t)$ that can effectively average out $H_{\text{SB}}$ and hence suppress decoherence.
 In particular we consider a continuous $H_c(t)$ of period $t_c$, whose time-ordered exponential defines a unitary operator $U_c(t)={\cal T}\exp[-i\int_0^{t}H_c(t')dt']$ ($\hbar=1$ throughout), with $U_c(t + t_c) = U_c(t)$.  The Magnus expansion \cite{Magnus1954,LidarPRA2005} indicates that
 if
\begin{equation}
\label{DDcondition2}
\int_0^{t_c} U_c^\dagger (t) H_{\text{SB}} U_c(t) \,dt = 0,
\end{equation}
then to its first order $H_{\text{SB}}$ is suppressed.  Extending previous studies for single-qubit and two-qubit systems \cite{FanchiniPRA20071,FanchiniPRA20072,ChaudhryPRA2012}, we choose
\begin{equation}
U_c(t) = e^{-2\pi i n_y J_y t/t_c} e^{-2\pi i n_x J_x t/t_c},
\label{Ueq}
\end{equation}
where $n_x$ and $n_y$ are non-zero integers. For any $n_x \neq n_y$, $U_c(t)$ in Eq.~(\ref{Ueq}) satisfies the first-order DD condition
of Eq.~(\ref{DDcondition2}).  Qualitatively, such $U_c(t)$ causes the collective angular momentum operators to rapidly rotate in
two independent directions and as a result, $H_{\text{SB}}$ is averaged out to zero. Using $i\, dU_c(t)/dt= H_c(t) U_c(t)$,
we obtain the following DD control Hamiltonian,
\begin{equation}
H_c(t) = \omega n_y J_y + \omega n_x [J_x \cos (\omega n_y t) - J_z \sin(\omega n_y t) ],
\label{Hceq}
\end{equation}
where $\omega\equiv 2\pi/t_c$.
The total system Hamiltonian $H_s(t) = H_{\text{OAT}} + H_c(t)$ then becomes
\begin{equation*}
H_s(t) = \chi J_x^2 + \omega n_y J_y + \omega n_x [J_x \sin (\omega n_y t) - J_z \cos (\omega n_y t)].
\label{tHamil}
\end{equation*}
To elaborate how $H_s(t)$ can be realized, we rotate the coordinate system along the $y$-axis by $\pi/2$,
transforming $H_s(t)$ to $H_s'(t)= \chi J_z^2 + \omega n_y J_y - \omega n_x [J_x \sin (\omega n_y t) + J_z \cos (\omega n_y t)]$.  We now comment on
each term of $H_s'(t)$.
The first $J_z^2$ term describes spin-spin interaction, as is realized in experiments \cite{GrossNature2010,RiedelNature2010}.
The last term linear in $J_z$ can be realized by an oscillating energy bias using for example a time-dependent Zeeman shift.
The $J_x$ and $J_y$ terms can be generated by use of electric-dipole interaction - considering a circularly polarized transition, a constant electric field along $y$ direction and an oscillating field along $x$ direction lead to the desired $J_x$ and $J_y$ terms \cite{Harochebook}.

With a continuous control Hamiltonian $H_c(t)$ implemented, decoherence can be well suppressed for sufficiently large $\omega$.
Two observations are in order. First, as shown in Eq.~(\ref{Hceq}), infinite DD solutions with different ($n_x$,$n_y$) combinations are found.
Second, the control Hamiltonian averages out $H_{\text{SB}}$ via fast modulations of $J_k$, so the system's self-interaction term $J_x^2$ is necessarily
modulated at the same time.  One opportunity is then emerging: among all the DD solutions, can we identify a particular type that modulate the
system's self-interaction Hamiltonian in a useful manner so as to enhance squeezing while suppressing decoherence?

 With the system decoupled from its environment, it can be shown that the system evolution operator is given by  $U_s(t)\approx U_c(t) e^{-i\bar{H} t}$ \cite{LidarPRA2005}, where the time-averaged Hamiltonian $\bar{H}$ is found to be
\begin{equation}
\bar{H} = \frac{\chi}{t_c} \int_0^{t_c} U_c^\dagger (t) J_x^2 U_c(t) \; dt.
\end{equation}
A straightforward though rather tedious calculation yields
\begin{eqnarray}
& & U_c^\dagger (t) J_x^2 U_c(t)  \nonumber  \\
&=& \frac{1}{2} \sin(2\omega n_x t) \sin^2 (\omega n_y t) [ J_z J_y + J_y J_z] \nonumber \\
&+& \frac{1}{2} \sin(2\omega n_y t) \cos (\omega n_x t) [J_x J_z + J_z J_x] \nonumber \\
&+& \frac{1}{2} \sin(2\omega n_y t) \sin (\omega n_x t) [J_x J_y + J_y J_x] \nonumber \\
&+& J_x^2 \cos^2(\omega n_y t) + J_y^2 \sin^2(\omega n_x t)\sin^2 (\omega n_y t) \nonumber \\
&+& J_z^2 \cos^2(\omega n_x t)\sin^2 (\omega n_y t).
\label{tedious}
\end{eqnarray}
\begin{figure}[t]
   \includegraphics{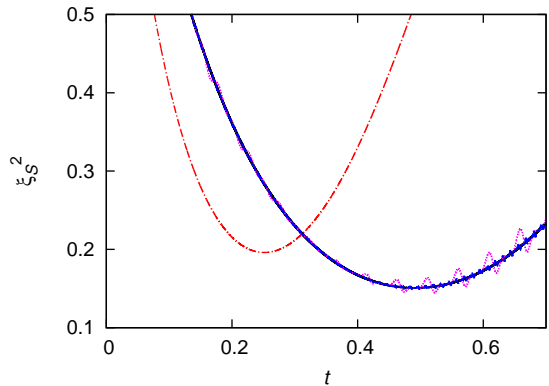}
   \centering
  	\caption{(color online) Spin squeezing measure
  $\xi_S^2$ against time $t$ using the OAT Hamiltonian (dot-dashed, red), $\bar{H}_{\text{DR}}$ (solid, dark blue), and $H_s(t)$ with $N_{\text{cyc}} = 5$ (dotted, magenta) and $N_{\text{cyc}} = 20$ (dashed, blue) for $N = 10$ (i.e., $J = 5$). Note that the dynamics generated by $H_s(t)$ with $N_{\text{cyc}} = 20$ are almost indistinguishable from the dynamics generated by $\bar{H}_{\text{DR}}$. Here we use $n_x = 2$ and $n_y = 1$, and $t_{\text{min}}$ was found to be approximately $0.491$.}
  	\label{J5DR}
\end{figure}

\begin{figure}[b]
   \includegraphics{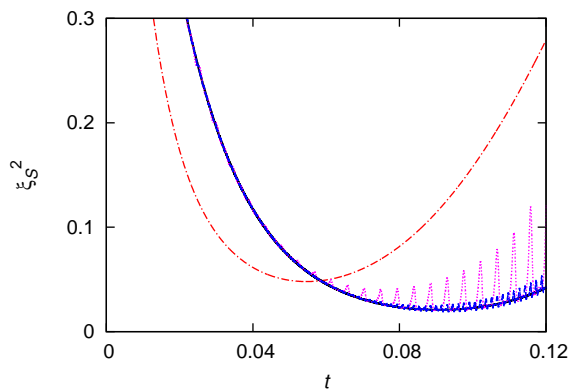}
   \centering
  	\caption{(color online) Same as in Fig.~\ref{J5DR}, but this time we have $N = 100$ (i.e., $J = 50$) with $N_{\text{cyc}} = 10$ (dotted, magenta), $N_{\text{cyc}} = 30$ (dashed, blue), and $t_{\text{min}} \approx 0.0909$. It is also observed here that the dynamics generated by $H_s(t)$ with $N_{\text{cyc}} = 30$ are well captured by
  the dynamics generated by $\bar{H}_{\text{DR}}$ (solid, dark blue).}
  	\label{J50DR}
\end{figure}
 Using Eq.~(\ref{tedious}), one finds the time-averaged Hamiltonian $\bar{H}$ has two different forms.
 Specifically, if $n_x \neq 2 n_y$, $\bar{H} = \frac{\chi}{4} J_x^2$ (up to a constant),
which is just the original OAT Hamiltonian with the nonlinear coefficient scaled down by a factor of four.
If $n_x = 2n_y$, which we call the ``double-resonance" (DR) condition,
we obtain (up to a constant)
\begin{equation}
\bar{H} = \bar{H}_{\text{DR}}= \frac{\chi}{4} \left( J_x^2 + J_x J_y + J_y J_x \right).
\end{equation}
Remarkably, $\bar{H}_{\text{DR}}$ is seen to be a mixture of a OAT Hamiltonian and a well-known two-axis twisting (TAT) Hamiltonian
$H_{\text{TAT}}=\chi(J_x J_y + J_y J_x)$ \cite{KitagawaUedaPRA1993, JianMaPhysicsReports2011}.  Since $H_{\text{TAT}}$ is known
to produce the best scaling of squeezing, we are motivated to examine the squeezing performance of $\bar{H}_{\text{DR}}$,
naturally obtained by one type of DD fields to fight against both relaxation and dephasing.


We consider an initial state $\ket{J,-J}$ describing all spins ``pointing down'', that is, an eigenstate of $J_z$ with eigenvalue $-N/2$.  To verify our expressions of $\bar{H}_{\text{DR}}$ and investigate its potential benefits
we first switch off $H_{\text{SB}}$ and evolve the wavefunction numerically
under $H_s(t) = H_{\text{OAT}} + H_c(t)$. We use $\xi_S^2 \equiv \frac{4 \; \text{min}(\Delta J_{\vec{n}_{\perp}})^2}{N}$ \cite{KitagawaUedaPRA1993}
to quantify squeezing, where $\vec{n}_{\perp}$ denotes a direction perpendicular to the mean spin direction and
the minimum is taken over all such directions.  The dynamical behavior of $\xi_S^2$ is found to be essentially the same as $\xi_R^2$
($\xi_R^2 =\xi_S^2 ({J}/|\langle \vec{J} \rangle|)^2  $), so only the behavior of  $\xi_S^2$ is presented below.
We set $\chi = 1$ and $t_c = t_{\text{min}}/N_{\text{cyc}}$, where $t_{\text{min}}$ is the optimal squeezing time for $\bar{H}_{\text{DR}}$,
and $N_{\text{cyc}}$ is a positive integer. In Fig.~\ref{J5DR}, the squeezing performance of $H_s(t)$ is compared with that of $H_{\text{OAT}} = \chi J_x^2$,
for $N = 10$.  It is seen that $H_s(t)$ generates much better squeezing - the minimum value of $\xi_S^2$ using $H_s(t)$ is approximately $0.15$, but we can only achieve a value of approximately $0.2$ using $H_{\text{OAT}}$. In addition,
it is seen from Fig.~\ref{J5DR} that the dynamics under $H_s(t)$ is indeed well captured by the dynamics under the time-averaged Hamiltonian
$\bar{H}_{\text{DR}}$. In particular,  within each period of $t_c$, $H_s(t)$ and $\bar{H}_{\text{DR}}$ yield some fluctuating differences in $\xi_S^2$, but at integer multiples of $t_c$ when $U_c(t)=1$, excellent agreement between them is observed. Further, if we reduce $t_c$ by increasing
the strength and the frequency of the periodic control Hamiltonian $H_c(t)$, the fluctuating differences become smaller. This reflects the fact that
$\bar{H}_{\text{DR}}$ is obtained under a first-order approximation.   We next increase $N$ tenfold and
essentially the same results are obtained in Fig.~\ref{J50DR},
but with the advantage gained by our DD fields displayed even more evidently. For example, in Fig.~2 it is seen that
the minimum values of $\xi_S^2$ using $H_s(t)$ and $H_{\text{OAT}}$ are approximately $0.019$ and $0.048$, respectively - so
more than a two-fold improvement can therefore be obtained.
Other calculations indicate that for larger $N$, the accuracy $\bar{H}_{\text{DR}}$ goes down with fixed $t_c$.
Thus,  a higher driving frequency $\omega$ would be more favored as $N$ increases.

\begin{figure}[t]
   \includegraphics{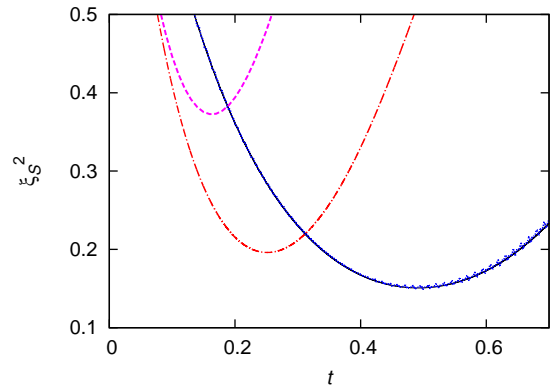}
   \centering
  	\caption{(color online) For $J = 5$, spin squeezing measure
  $\xi_S^2$ against time $t$ using the bare OAT Hamiltonian without noise (dot-dashed, red), $\bar{H}_{\text{DR}}$ (solid, dark blue), the OAT Hamiltonian with noise but without DD fields (dashed, magenta), and the OAT Hamiltonian in the presence noise and the DD fields with $n_x = 2$, $n_y = 1$, $t_{\text{min}} \approx 0.491$, and $N_{\text{cyc}} = 20$ (dotted blue line, which is almost on top of the solid line). An average over $2000$ sample paths of the noise was taken. The noise parameters are $\alpha = 2$ and $\sigma^2 = 20$.}
  	\label{NoiseJ5DR}
\end{figure}

\begin{figure}[b]
   \includegraphics{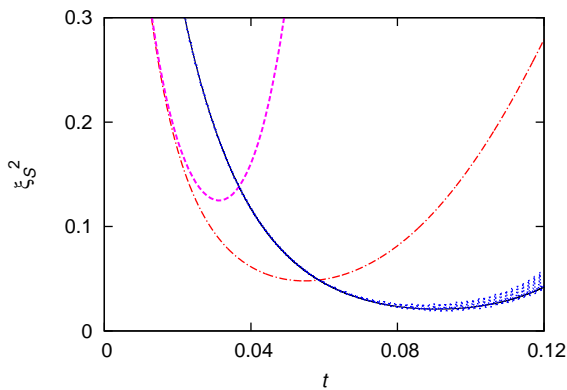}
   \centering
  	\caption{(color online) Same as in Fig.~\ref{NoiseJ5DR}, but now with $J = 50$, $t_{\text{min}} \approx 0.0909$, $N_{\text{cyc}} = 30$, and $\sigma^2 = 100$. An average over $100$ sample paths of the noise was taken.  Similar to what is observed in Fig.~\ref{NoiseJ5DR}, the squeezing performance of the OAT Hamiltonian in the presence of noise and DD fields (dotted, blue)
  is much better than that of a bare OAT Hamiltonian in the absence of noise (dot-dashed, red).}
  	\label{NoiseJ50DR}
\end{figure}

Having confirmed that the double-resonance condition $n_x=2n_y$ is useful for spin squeezing,
let us now turn to the full problem by switching on the system-environment coupling.
For convenience we model $B_x$, $B_y$ and $B_z$ in Eq.~\eqref{systemenvinteraction} as three independent Gaussian colored noise processes,
with the same inverse correlation time $\alpha$ and noise variance $\sigma^2$.
We numerically compute the dynamics of squeezing for $H_s(t)$ in the presence of noise and then compare it with that generated by $H_{\text{OAT}}$, with noise or without noise. As shown in Figs.~\ref{NoiseJ5DR} and \ref{NoiseJ50DR}, $H_s(t)$ with noise yields much better squeezing than
$H_{\text{OAT}}$ with noise.  This may be understood as an outcome of decoherence suppression.  On the other hand,
 $H_s(t)$ with noise also generates better squeezing than
$H_{\text{OAT}}$ in the absence of noise.  Hence our DD fields have played one more role in addition to decoherence suppression.
Note also that the time to obtain maximum squeezing is in excellent agreement with that obtained from $\bar{H}_{\text{DR}}$,
hence avoiding decoherence effects on the optimal squeezing time and also confirming again the usefulness of $\bar{H}_{\text{DR}}$
in predicting the optimal squeezing time.

\begin{figure}[h]
   \includegraphics{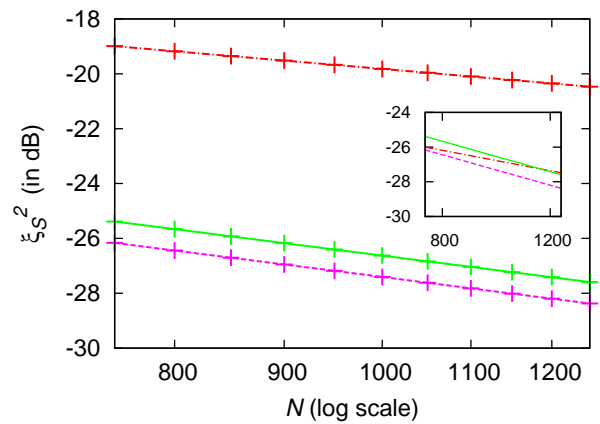}
   \centering
  	\caption{(color online) Minimum value of $\xi_S^2$ - in dB, defined as $10 \, \text{log}(\xi_S^2)$ - plotted against $N$ (log scale) for dynamics generated by the OAT Hamiltonian (upper, dot-dashed, red), the TAT Hamiltonian (lower, dashed, magenta), and the time-averaged Hamiltonian $\bar{H}_{\text{DR}}$ (middle, solid, green). In the inset, the upper (dot-dashed, red) line has been transported vertically to show clearly the better scaling behavior of $\bar{H}_{\text{DR}}$ and the TAT Hamiltonian. The behavior of $\xi_R^2$ (not shown) is similar.}
  	\label{sqdecibels}
\end{figure}

  Having shown how DD fields may suppress decoherence and enhance the spin
squeezing generation as a unitary process, we finally investigate
how the squeezing performance of $\bar{H}_{\text{DR}}$ scales with $N$. Within the validity regime of $\bar{H}_{\text{DR}}$
as an effective Hamiltonian (for describing the dynamics associated with the OAT Hamiltonian in the presence of noise and continuous DD fields),
the scaling of the squeezing performance of $\bar{H}_{\text{DR}}$ with $N$
represents to what degree our
DD fields can protect and enhance spin squeezing.  Calculations for even larger values of $N$ then become necessary.
We first compare the performance of $\bar{H}_{\text{DR}}$ with what is known to give the best scaling behavior, namely,
the TAT Hamiltonian $H_{\text{TAT}}=\chi (J_x J_y+ J_y J_x)$.  Significantly, although $\bar{H}_{\text{DR}}$ produces slightly
less squeezing than $H_{\text{TAT}}$, two close and parallel lines describing their respective performance
are seen in Fig.~\ref{sqdecibels}, indicating that both
cases give the $\xi_{S}^2 \sim 1/N$ scaling. By contrast, Fig.~\ref{sqdecibels} also presents the $1/N^{2/3}$ scaling of
the OAT Hamiltonian [also see the inset of Fig.~\ref{sqdecibels} for a comparison of two different scalings].
The DD fields under the double-resonance condition
hence allows squeezing to occur in the presence of noise and in the mean time
brings about a squeezing enhancement factor of $N^{1/3}$, which is in principle unlimited as $N$ increases.

It is also interesting to note how this work differs from a recent proposal for realizing TAT Hamiltonian by applying a designed pulse sequence to a OAT Hamiltonian~\cite{LiuPRL2011}.  While our starting point is continuous DD fields for decoherence suppression, the short
control pulses considered in Ref.~\cite{LiuPRL2011}
do not average out $H_{\text{SB}}$ in Eq.~(1) to zero.  Further, the effective Hamiltonian $\bar{H}_{\text{DR}}$ found here under a double-resonance condition
is a mixture of OAT and TAT Hamiltonians. To our knowledge, $\bar{H}_{\text{DR}}$ is a newly found, physically motivated Hamiltonian that
can generate the $\xi_{S}^2 \sim 1/N$ scaling.

To conclude, by considering a class of continuous fields to suppress both dephasing and relaxation in the dynamics of
spin squeezing, we are able to identify a special type of DD solutions that can effectively
yield a previously unknown spin squeezing Hamiltonian, generating the $1/N$ scaling of squeezing performance in the presence of an environment.
With their dual roles in decoherence suppression and in generating more useful
quantum evolution identified, the found DD fields are appealing from an experimental point of view.
Our results should be able to help design new experimental studies of spin squeezing based on one-axis twisting Hamiltonians (such as those using two-mode BEC).
Indeed, by exploiting system's own spin-spin interaction
Hamiltonian under the modulation of continuous DD fields, we expect to see other interesting DD designs
that can carry out desired quantum operations while protecting quantum coherence.

A.Z.C. would like to thank Derek Ho and Juzar Thingna for helpful discussions.


\begin{thebibliography}{99}
\bibitem{footnoteprecision} In the prototype Ramsey interferometry setup, the phase uncertainty using uncorrelated particles is $\Delta \phi = 1/\sqrt{N}$, while using a squeezed state of $N$ particles, the uncertainty is $\Delta \phi = \xi_R/\sqrt{N}$, where $\xi_R$ is a measure of spin squeezing.
\end{thebibliography}
\end{document}